\documentclass[preprint]{revtex4}%
\usepackage{graphicx}
\usepackage{dcolumn}
\usepackage{bm}
\usepackage{amsmath}
\usepackage{amsfonts}
\usepackage{amssymb}%
\begin{document}
\title{Calculations of Hubbard U from first-principles.\textbf{ }}
\author{F. Aryasetiawan$^{1,2}$, K. Karlsson$^{3,4}$, O. Jepsen$^{4}$ and U.
Sch\"{o}nberger$^{4}$,}
\affiliation{$^{1}$\textit{Research Institute for Computational Sciences, AIST, }}
\affiliation{\textit{Umezono 1-1-1, Tsukuba Central 2, Tsukuba Ibaraki 305-8568 Japan,}}
\affiliation{$^{2}$CREST, Japan Science and Technology Agency}
\affiliation{$^{3}$\textit{Department of Natural Science,}}
\affiliation{\textit{H\"{o}gskolan i Sk\"{o}vde, 54128 Sk\"{o}vde, Sweden}}
\affiliation{$^{4}$\textit{Max Planck Institut f\"{u}r Festk\"{o}rperforschung}}
\affiliation{\textit{D-705 06 Stuttgart, Germany} }
\date{}

\begin{abstract}
The Hubbard \emph{U} of the \emph{3d} transition metal series as well as
SrVO$_{3}$, YTiO$_{3}$, Ce and Gd has been estimated using a recently proposed
scheme based on the random-phase approximation. The values obtained are
generally in good accord with the values often used in model calculations but
for some cases the estimated values are somewhat smaller than those used in
the literature. We have also calculated the frequency-dependent \emph{U} for
some of the materials. The strong frequency dependence of \emph{U} in some of
the cases considered in this paper suggests that the static value of \emph{U}
may not be the most appropriate one to use in model calculations. We have also
made comparison with the constrained LDA method and found some discrepancies
in a number of cases. We emphasize that our scheme and the constrained LDA
method theoretically ought to give similar results and the discrepancies may
be attributed to technical difficulties in performing calculations based on
currently implemented constrained LDA schemes.

\end{abstract}
\maketitle

\section{Introduction}

Many-electron problem in real materials is too complicated to be tackled
directly from first principles. Direct conventional methods for calculating
excited-state properties are available and one of the most successful among
these is the GW approximation (GWA)\cite{hedin,ferdi}. However, it has proven
difficult to go beyond the GWA purely from first principles for a number of
reasons. Seen from diagrammatic point of view, the GWA is a sum of bubble
diagrams corresponding to the random-phase approximation (RPA)\cite{pines},
which can be more generally regarded as time-dependent Hartree approximation.
Firstly it is far from obvious which classes of diagrams should be included
beyond the RPA. Secondly when we do include another class of diagrams, the
lowest order diagram, i.e., second order, is likely to be included twice. This
is because there are only very few lower order diagrams so that different
classes of diagrams will tend to have an overlap of the lowest-order diagrams.
This double counting has to be taken into account. However, for an arbitrary
frequency, the imaginary part of the second-order self-energy may be larger
than, e.g., the imaginary part of the GW self-energy. When the second-order
self-energy is subtracted for double counting this may lead to a wrong sign of
the imaginary part of the self-energy with unphysical consequences such as a
negative spectral function.

It is therefore worthwhile to consider an alternative approach of tackling the
many-electron problem in real materials. We note that of the large number of
degrees of freedom only a few are actually relevant for electron correlations.
A typical problem that is difficult to treat with conventional methods
corresponds to a system with localized orbitals embedded in extended states.
Many so-called strongly correlated materials, such as perovskites, are of this
type, consisting of localized \emph{3d} or \emph{4f} orbitals embedded in
extended \emph{s-p} states. Electrons living in the localized orbitals
experience strong correlations among each other with a subtle coupling to the
extended states resulting in a complicated many-electron problem. The RPA is
probably insufficient for treating correlations among electrons in localized
orbitals. One may speculate that many classes of diagrams contribute with
similar weight to correlations, making it very difficult to go beyond the RPA
by including a few extra classes of diagrams.

The idea of isolating a few degrees of freedom relevant for correlation has
been utilized for many years in the Hubbard model or the Anderson impurity
model. Renormalized or screened Coulomb interaction (Hubbard \emph{U}) is kept
among electrons living in the localized orbitals. In the Anderson model some
states coupled to the localized orbitals are also kept but without Coulomb
interaction and the rest of the states are downfolded resulting in a
renormalized Coulomb interaction. Although the physical picture seems clear it
is not immediately evident how to calculate the Hubbard \emph{U}. Indeed in
practice it has usually been regarded as an adjustable parameter. Adjustable
parameters are unsatisfactory since they limit the quantitative predictive
power of the model. Even qualitatively adjustable parameters might lead to
misleading conclusions. It is therefore highly desirable to find a systematic
way of calculating the parameters in the Hubbard model, in particular the
Hubbard \emph{U}, from realistic electronic structure calculations.

The problem of determining the Hubbard \emph{U} from first principles has been
addressed by a number of authors. One of the earliest works is the constrained
local density approximation (cLDA) approach \cite{olle1,hubert,mcmahan} where
the Hubbard \emph{U} is calculated from the total energy variation with
respect to the occupation number of the localized orbitals. A further
improvement of this scheme was recently proposed \cite{imada,stefano,nakamura}%
. A different approach based on the random-phase approximation (RPA) was later
introduced \cite{springer}, which allows for the calculations of the matrix
elements of the Hubbard \emph{U} and its energy dependence. More recently, it
was shown that it is possible to calculate the Hubbard \emph{U} systematically
from first principles \cite{ferdi1}. It was soon realized that these two
methods of calculating the Hubbard \emph{U}, RPA and cLDA, do not yield the
same results. This is very puzzling since the two approaches are supposed to
give the effective interaction of the localized electrons and ought to give
the same results.

The purpose of the present work is to present results for the Hubbard
\emph{U}\textbf{ }of the \emph{3d} series and a few other materials (Ce,
SrVO$_{3}$, YTiO$_{3})$ and to analyze the origin of the discrepancy between
RPA and cLDA results.

\section{Theory}

\subsection{Constrained RPA}

The fully screened Coulomb interaction is given by%

\begin{equation}
W=[1-vP]^{-1}v
\end{equation}
where \emph{v} is the bare Coulomb interaction and \emph{P }is the
non-interacting polarization given by
\begin{align}
P(\mathbf{r,r}^{\prime};\omega)  &  =\sum_{i}^{occ}\sum_{j}^{unocc}\psi
_{i}(\mathbf{r)}\psi_{i}^{\ast}(\mathbf{r}^{\prime})\psi_{j}^{\ast
}(\mathbf{r)}\psi_{j}(\mathbf{r}^{\prime})\nonumber\\
&  \times\left\{  \frac{1}{\omega-\varepsilon_{j}+\varepsilon_{i}+i0^{+}%
}-\frac{1}{\omega+\varepsilon_{j}-\varepsilon_{i}-i0^{+}}\right\}  . \label{P}%
\end{align}
where $\{\psi_{i},\varepsilon_{i}\}$ are one-particle Bloch eigenfunctions and
eigenvalues corresponding to the system's band structure. For systems with a
narrow \emph{3d} or \emph{4f} band across the Fermi level, typical of strongly
correlated materials, we may divide the polarization into $P=P_{d}+P_{r}$, in
which $P_{d}$ includes only \emph{3d} to \emph{3d} transitions (i.e., limiting
the summations in (\ref{P}) to $i,j\in\{\psi_{d}\}$), and $P_{r}$ be the rest
of the polarization. \ In a previous publication \cite{ferdi1} it was shown
that the following quantity can be interpreted as the effective interaction
among electrons living in the narrow band (Hubbard \emph{U)}:
\begin{equation}
U(\omega)=[1-vP_{r}(\omega)]^{-1}v \label{Wr}%
\end{equation}
where $U$ can be related to the fully screened interaction \emph{W} by the
following \emph{identity:}
\begin{equation}
W=[1-UP_{d}]^{-1}U. \label{W}%
\end{equation}
This identity explicitly shows that the interaction between the \emph{3d}
electrons is given by a frequency-dependent interaction $U$. Thus the
remaining screening channels in the Hubbard model associated with the
\emph{3d} electrons, represented by the \emph{3d}-\emph{3d} polarization
$P_{d},$ further screen $U$ to give the fully screened interaction $W$. In
analogy to the constrained LDA\ method, we refer the method of calculating the
Hubbard \emph{U} as in (\ref{Wr}) as \textquotedblright constrained
RPA\textquotedblright\ (cRPA) because we have constrained the polarization to
exclude \emph{3d}-\emph{3d} transitions. In contrast to Ref. \cite{stefano},
it is not necessary to subtract the contribution arising from rehybridization
of the non-interacting Kohn-Sham band structure because the wave functions and
eigenvalues appearing in Eq. (\ref{P}) are fixed.

It is noteworthy that \emph{U} in Eq. (\ref{Wr}) is a function of positions
\textbf{r} and \textbf{r'}, independent of basis functions. This is because
the polarization in Eq. (\ref{P}) depends on the Bloch wave functions and
eigenvalues, which can be calculated in any basis. Thus, our method is
basis-independent. In the following, we retain only the local components of
the effective interaction on the same atomic site by taking the following
matrix element:%

\begin{equation}
U=\int d^{3}rd^{3}r^{\prime}\ |\phi_{3d}(\mathbf{r})|^{2}U(\mathbf{r,r}%
^{\prime})|\phi_{3d}(\mathbf{r}^{\prime})|^{2} \label{U}%
\end{equation}
where $\phi_{3d}$ is a \emph{3d} LMTO \cite{andersen-lmto} orbital centered on
an atomic site and the interaction $U(\mathbf{r,r}^{\prime})$ is the static
($\omega=0)$ value of Eq. (\ref{Wr}). In calculating \emph{U} we have
approximated $\phi_{3d}$ by the "head" of the LMTO, i.e., the solution to the
Schr\"{o}dinger equation inside the atomic sphere. This is expected to be a
reasonable approximation because the $\emph{3d}$ states are rather localized.
The lattice Hubbard model with a \textit{static} interaction \emph{U }is given
by
\begin{equation}
H=\sum_{Rn,R^{\prime}n^{\prime}}c_{Rn}^{\dagger}h_{Rn,R^{\prime}n^{\prime}%
}c_{R^{\prime}n^{\prime}}+\frac{1}{2}\sum_{R,nn^{\prime},mm^{\prime}}%
c_{Rn}^{\dagger}c_{Rn^{\prime}}U_{nn^{\prime},mm^{\prime}}c_{Rm}^{\dagger
}c_{Rm^{\prime}} \label{ham}%
\end{equation}
$h_{Rn,R^{\prime}n^{\prime}}$ is the one particle part of the Hamiltonian
consisting of hopping and the orbital energy. A model with a
frequency-dependent \emph{U} can be formulated within the action integral
approach but cannot be formulated within the Hamiltonian approach
\cite{ferdi1}. It is clear that the \emph{U} entering the Hubbard model will
inevitably depend on the choice of the one-particle basis $\phi_{3d}$ defining
the annihilation and creation operators, no matter what method we use to
calculate $U(\mathbf{r,r}^{\prime})$. LMTO is just one possible choice for the
one-particle orbitals but other choices are perfectly legitimate. For example,
the newly developed NMTO (N corresponds to the order of the Taylor expansion
in energy of the orbital) \cite{nmto} and the recently proposed maximally
localized Wannier orbitals \cite{vanderbilt} are possible choices.

\subsection{Constrained LDA}

The derivative of the total energy with respect to an occupation number
$n_{i}$ of a given state can be related to its Kohn-Sham eigenvalue as follows
\cite{janak}:%

\begin{equation}
\frac{\partial E}{\partial n_{i}}=\varepsilon_{i}. \label{janak}%
\end{equation}
From the Kohn-Sham equation \cite{kohn} one can show that following relation
holds \cite{vonbarth, springer}%

\begin{equation}
\frac{\partial\varepsilon_{i}}{\partial n_{j}}=\frac{\partial^{2}E}{\partial
n_{i}\partial n_{j}}=\left\langle ij|(v+f_{xc})\epsilon^{-1}|ij\right\rangle
\label{d2E}%
\end{equation}
where%

\[
f_{xc}(\mathbf{r,r}^{\prime})=\frac{\partial^{2}E_{xc}}{\partial
\rho(\mathbf{r)}\partial\rho(\mathbf{r}^{\prime})}=\frac{\partial
v_{xc}(\mathbf{r)}}{\partial\rho(\mathbf{r}^{\prime})}%
\]
and $\epsilon^{-1}$ is the inverse dielectric function which can be expressed
in terms of the linear density-density response function $R=\delta\rho/\delta
v_{ext},$ i.e., the change in density $\rho$ with respect to an external
perturbation $v_{ext}$, as%

\[
\epsilon^{-1}=1+Rv.
\]
The integral in (\ref{d2E}) is defined as%

\begin{equation}
\left\langle ij|F|ij\right\rangle =\int d^{3}rd^{3}r^{\prime}\ |\phi
_{i}(\mathbf{r)|}^{2}F(\mathbf{r,r}^{\prime})|\phi_{j}(\mathbf{r}^{\prime
}\mathbf{)|}^{2}.
\end{equation}
Within the RPA, which is equivalent to time-dependent Hartree approximation,
$f_{xc}=0$ implying that $(v+f_{xc})\epsilon^{-1}=v+vRv=W,$ i.e., the change
in the Kohn-Sham eigenvalue is directly related to the screened interaction.
In fact in practice $f_{xc}\ll v$ so that to a good approximation we may
assume that $\partial\varepsilon_{i}/\partial n_{j}=\left\langle
ij|W|ij\right\rangle .$

The idea in a constrained LDA calculation (cLDA) is to perform a
self-consistent (supercell) total energy calculation with a constrained
\emph{3d} occupancy on the so called impurity atom. Furthermore, the coupling
between the impurity (\emph{3d}-level) and the rest of the system, which is
explicitly included in the model Hamiltonian, is removed in order to avoid
double counting. According to (\ref{d2E}) the Hubbard $U$ is then given by the
change in the \emph{3d}-level when the number of localized \emph{3d} electrons
are varied and the hopping for the localized \emph{3d} electrons is
suppressed:
\begin{equation}
U=\frac{\partial C_{3d}}{\partial n_{3d}}, \label{C}%
\end{equation}
where $C_{3d}$ is the center of the \emph{3d} band. By suppressing the hopping
of the \emph{3d} electrons, the screened interaction so obtained corresponds
to an effective interaction without the participation of screening from the
\emph{3d} electrons, which is what we mean by the Hubbard \emph{U}. However,
by cutting off the hopping of the \emph{3d} electrons, we do not only
eliminate hopping to \emph{3d} bands but also hopping to bands other than
\emph{3d}. We will elucidate later that the neglect of the latter hopping is
the main origin of the discrepancy between cLDA\ and cRPA.

Within the LMTO-ASA (Atomic Sphere Approximation) scheme, the \emph{3d} level,
or in fact the band-center $C_{3d}$, is obtained by solving the radial
Schr\"{o}dinger equation for the impurity atom using fix boundary condition
(BC) at the muffin-tin (MT) sphere \cite{andersen-lmto}. The BC are
arbitrarily chosen so the radial wavefunction in the sphere $\phi_{3d}$,
matches smoothly to $r^{-l-1}$ at the MT sphere, which corresponds to a
logarithmic derivative for the \emph{3d}-states of $D_{l}=-l-1=-3$. We note
that in the above definition $\phi_{3d}$ is allowed to relax
(self-consistently) upon a change in the \emph{3d} charge. A decrease in the
number of \emph{3d} electrons will make the potential more attractive and
consequently the wavefunction will contract. This effect is compensated due to
remaining electrons located outside the impurity sphere, which tend to
screen.
In addition we have performed so called modified $U$ calculations using a
$\mathit{fixed}$ wavefunction, obtained in an ordinary bulk calculation, in
order to make sure that the wavefunction $\phi_{3d}$ coincides with the one
used in the cRPA approach. Then the definition of $U$ has to be slightly
modified (denoted cLDA (modified) in Fig. (\ref{urpa})), and defined as the
change in the expectation value of the impurity potential related to changes
in \emph{3d} occupancy:
\begin{equation}
U=\frac{\partial\langle\phi_{3d}\mid V\mid\phi_{3d}\rangle}{\partial n_{3d}%
}=\frac{\partial}{\partial n_{3d}}\int d^{3}r\phi_{3d}^{2}(r)V(r).
\label{mod1}%
\end{equation}
The above formula is obtained by observing that to first order in the
perturbing potential the change in $\varepsilon_{3d}$ or $C_{3d}$ is given by
$\langle\phi_{3d}\mid V\mid\phi_{3d}\rangle$. In order to validate such a
procedure for evaluating $U$, we have also adopted a scheme where firstly the
supercell calculation is done until self-consistency using the fixed bulk wave
functions, i.e., we perform a constrained calculation as usual but with fixed
wave functions. Secondly the resulting self-consistent potential is used only
\textit{once} (BC $D_{l}=-3$) to obtain the $C_{3d}$ level by solving the
radial Schr\"{o}dinger equation. Our results confirm that the value of $U$
obtained from Eq. (\ref{C}) using this one-shot iteration for the $C_{3d}$
level is in fact the same as the value obtained from the modified definition
in Eq. (\ref{mod1}).

We must have in mind that in the LMTO-ASA method \cite{andersen-lmto}, any
polarization of the screening charge is neglected. However, the screening
charge inside the atomic sphere (on-site screening) is taken care of with good
accuracy, but the non-local screening from other spheres is merely due to
point-charges located at the sphere-centers (Madelung screening). In reality,
this charge is expected to be located somewhat closer to the impurity sphere,
thus reducing the value of the calculated $U$. As a consequence systems with
almost all screening charge inside the impurity sphere are expected to be well described.

The beauty of the LMTO method is that the basis used in the band structure
calculation is the same as the one-particle basis defining the annihilation
and creation operator of the Hubbard model. Thus by constraining the
occupation number of the orbital one directly obtains the corresponding
\emph{U }from the change in the orbital energy with respect to the occupation
number as in Eq. (\ref{d2E}) or (\ref{C}). This is in contrast to methods not
based on localized orbitals, \cite{stefano, nakamura} where projection to some
localized orbitals defining to the one-particle basis of the model Hamiltonian
is necessary.

\section{Results and Discussions}

\subsection{SrVO$_{3}$}%

\begin{figure}
[ptb]
\begin{center}
\includegraphics[
height=2.783in,
width=2.4025in
]%
{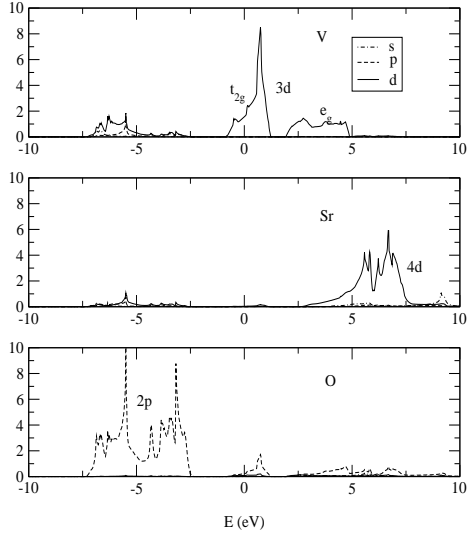}%
\caption{Partial density of states of SrVO$_{3}.$}%
\label{dossrvo3}%
\end{center}
\end{figure}
%

\begin{figure}
[ptb]
\begin{center}
\includegraphics[
height=2.6152in,
width=3.8354in
]%
{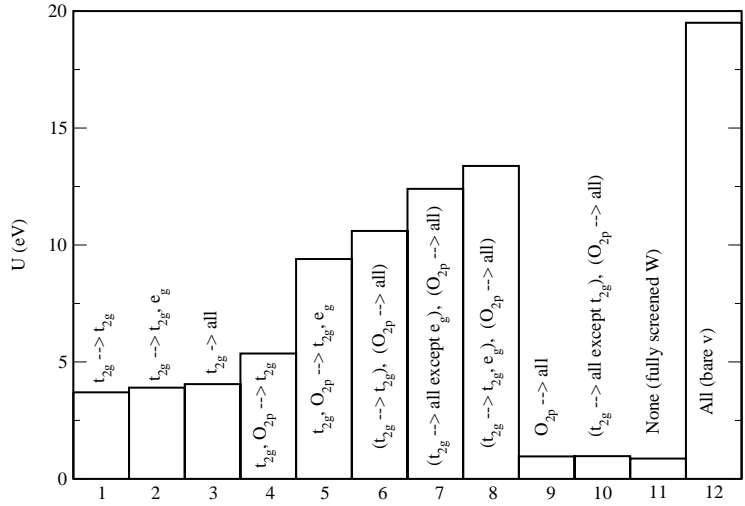}%
\caption{The Hubbard \emph{U} of SrVO$_{3}$ obtained by eliminating various
transitions as indicated in the picture. For example, case 5 corresponds to
eliminating transitions t$_{2g}$--$>$t$_{2g}$, t$_{2g}$--$>$e$_{g}$, O$_{2p}%
$--$>$t$_{2g}$, and O$_{2p}$--$>$e$_{g}$ and case 6 corresponds to eliminating
t$_{2g}$--$>$t$_{2g}$ transition and all transitions from O$_{2p}$. Discussion
of the result is described in the text.}%
\label{Ut2g}%
\end{center}
\end{figure}
%

\begin{figure}
[ptb]
\begin{center}
\includegraphics[
height=2.3163in,
width=3.1855in
]%
{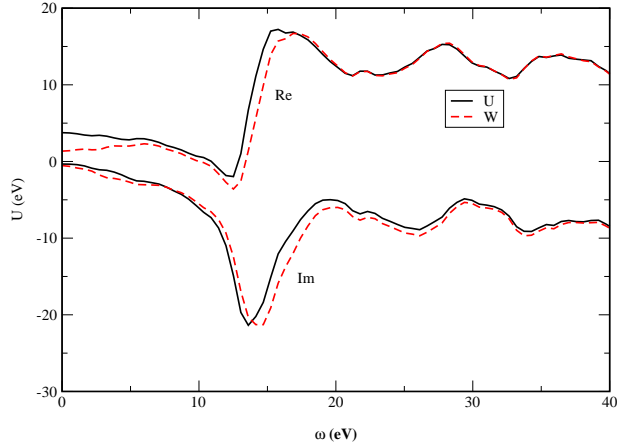}%
\caption{Frequency dependent \emph{U} and \emph{W} of SrVO$_{3}.$}%
\label{srvo3}%
\end{center}
\end{figure}

SrVO$_{3}$ is an ideal case because across the Fermi level there is a narrow
band of mainly t$_{2g}$ character originating from V \emph{3d}, well separated
from other bands. This allows for an unambiguous application of Eq.
(\ref{Wr}). To aid in visualizing the various transitions between occupied and
unoccupied states the partial density of states of SrVO$_{3}$ are displayed in
Fig. \ref{dossrvo3}. The effects of various possible screening channels on the
value of the screened interaction are calculated systematically and shown in
Fig. \ref{Ut2g}. From the figure we can read the value of \emph{U} for the
t$_{2g}$ band, namely, \emph{U}=3.5 eV. As can be seen from the figure,
eliminating transitions from t$_{2g}$ band to e$_{g}$ band has hardly any
effect on \emph{U}. This implies that the Hubbard \emph{U} for a model with
only t$_{2g}$ orbitals (case 1) is approximately the same as the one for a
model with both t$_{2g}$ and e$_{g}$ orbitals (case 2). In fact, eliminating
all transitions from the t$_{2g}$ band (case 3) has little effect on \emph{U}.
This is because the small number of \emph{3d} electrons ($\sim1)$ contribute
little to the polarization other than \emph{3d-3d }polarization. From cases 4,
5, and 6 it becomes evident that screening from the oxygen \emph{2p} electrons
is very important. In particular, the transitions from oxygen \emph{2p} to
$t_{2g}$ and $e_{g}$ bands are most significant (see also a recent work of
Solovyev and Imada \cite{imada}). However, cases 9 and 10 are surprising: when
all transitions from the oxygen \emph{2p} are eliminated, the resulting
screened interaction is almost the same as the fully screened interaction
\emph{W} (case 11). This means that the $t_{2g}$ electrons alone are very
efficient in performing the screening so that after the $t_{2g}$ electrons
have performed the screening there is little left for the oxygen \emph{2p}
electrons to screen. In retrospect, it is a reasonable result because t$_{2g}%
$--%
$>$%
t$_{2g}$ screening is metallic, which is very efficient. Similarly, when the
oxygen \emph{2p} electrons are allowed to screen first (case 3), there is
little left for the $t_{2g}$ electrons to screen. Thus, cases 3 and 9 are not contradictory.

It is interesting to compare the value of \emph{U}\textbf{ }calculated using
cRPA and cLDA. The value of \emph{U} for the t$_{2g\text{ }}$and e$_{g}$ bands
obtained from cLDA is 9.5 eV and the modified cLDA yields a value of 8.8 eV.
Since in the implementation of cLDA, hopping to the \emph{3d} orbitals on a
given impurity site in a supercell is prohibited, a fair comparison with cRPA
would be to eliminate transitions from and to the \emph{3d} orbitals (both
t$_{2g}$ and e$_{g}$). In this interpretation of cLDA, we should therefore
make comparison with case 5, corresponding to the case where transitions from
O$_{2p}$ to \emph{3d} bands (both \emph{t}$_{2g}$ and \emph{e}$_{g}$) are also
prohibited. Indeed the value corresponding to case 5 is very close to the
modified cLDA value. We should also eliminate the t$_{2g}$--%
$>$%
non-t$_{2g}$ screening but as shown by cases 1, 2, and 3 in Fig. (\ref{Ut2g}),
this screening channel is not important.

The frequency-dependent \emph{U} and \emph{W} are displayed in Fig.
(\ref{srvo3}). There is a qualitative difference in the frequency dependence
of the imaginary part of \emph{U }of SrVO$_{3}$ and late transition metals. In
the former the imaginary part is dominated by a single plasmon excitation
while the latter is characterized by a broad excitations with no distinct
plasmon excitation. This is a consequence of the small polarization
contribution of the \emph{t}$_{2g}$ electrons to non-\emph{3d} orbitals in the
former so that the plasmon excitation is dominated by the free-electron-like
O$_{2p}.$ In late transition metals, such as nickel, the \emph{3d} electrons
themselves contribute very significantly to screening in the form of
polarization to non-\emph{3d} orbitals. The localized non-free-electron-like
nature of the \emph{3d} electrons may be responsible for the broad excitation
energies observed in Im \emph{U }and Im \emph{W}.

We have also calculated \emph{U} for YTiO$_{3}$ with a result equal to 4.0 eV.
This value is significantly smaller than the value of 5.0 eV used in LDA+DMFT
(Dynamical Mean-Field Theory) calculations. \cite{pavarini} The value of 5.0
eV is needed in order to open up a gap when starting from a metallic state in
the LDA. While a smaller value of 3-4 eV would still be acceptable in the case
of SrVO$_{3}$ or CaVO$_{3}$ since they are both metals, such value of \emph{U}
would just barely open up a gap in the case of YTiO$_{3}$. This poses a
fundamental question concerning the value of the Hubbard \emph{U} that is
appropriate for DMFT. It could well be that due to the mapping to an impurity
model, DMFT requires a larger \emph{U} than the corresponding value for a
lattice model. However, this runs counter to our intuition because by mapping
to an impurity model one would expect that part of the screening process
arising from the impurity's neighboring sites should already be included in
the impurity \emph{U}, which should therefore be smaller than the lattice
\emph{U}. Another possibility is that the RPA is not sufficient for treating
local screening and more accurate approximation beyond the RPA may be needed.
This question is now under study. We have also performed cLDA calculations
giving a value of \emph{U}=7.5 eV and 6.3 eV with a modified cLDA. We have not
analyzed in detail the various transitions in YTiO$_{3}$ since we believe they
are essentially similar to those of SrVO$_{3}$.

\subsection{Cerium}%

\begin{figure}
[ptb]
\begin{center}
\includegraphics[
height=2.783in,
width=2.7614in
]%
{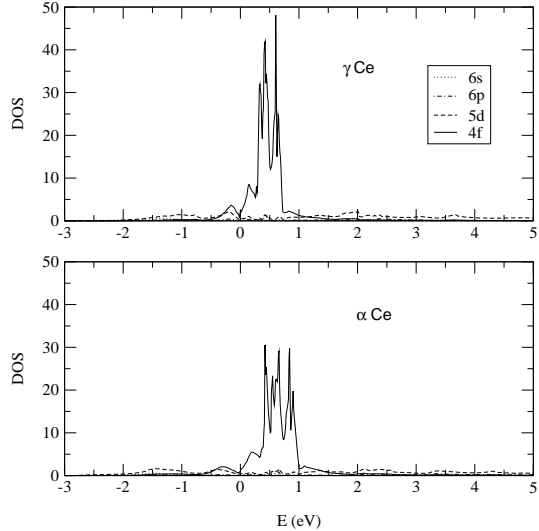}%
\caption{Partial density of states of $\alpha$Ce and $\gamma$Ce.}%
\label{dosce}%
\end{center}
\end{figure}

Another interesting case to consider is cerium although it is less ideal than
SrVO$_{3}$ due to a mixing between the \emph{4f} and \emph{5d} orbitals in the
valence states. The partial densities of states of $\alpha$Ce and $\gamma$Ce
are shown in Fig. \ref{dosce}. In Tables I and II the values of \emph{U} for a
number of energy windows are shown.

\begin{table}[ptb]%
\begin{tabular}
[c]{ll}%
Energy window (eV) & U (eV)\\
(-2.0, 1.5) & 7.9\\
(-1.5, 1.5) & 7.6\\
(-1.0, 1.5) & 5.7\\
(-0.5, 1.5) & 3.3\\
(-0.5, 1.0) & 2.9\\
(-0.5, 1.7) & 3.3\\
(-0.5, 2.0) & 3.4
\end{tabular}
\caption{\emph{U} of $\gamma$Ce as a function of energy window}%
\end{table}\ \ \begin{table}[ptb]%
\begin{tabular}
[c]{ll}%
Energy window (eV) & U (eV)\\
(-2.0, 1.7) & 6.6\\
(-1.5, 1.7) & 5.4\\
(-1.0, 1.7) & 4.3\\
(-0.7, 1.7) & 3.2\\
(-0.7, 2.0) & 3.3\\
(-0.7, 3.0) & 3.4\\
&
\end{tabular}
\caption{\emph{U} of $\alpha$Ce as a function of energy window}%
\end{table}

A reasonable choice of energy window for $\alpha$Ce and $\gamma$Ce are (-0.7,
1.7) and (-0.5, 1.5), respectively since these cover the \emph{4f} states.
This choice gives a value of about 3.2-3.3 eV. It is clear from the tables
that the value of \emph{U} is very sensitive to the choice of the lower energy
bound. Thus \emph{U} is more than doubled when the lower bound is taken to be
-2.0 eV, which corresponds to covering the entire occupied valence states. On
the other hand, due to the small number of \emph{4f} electrons, \emph{U} is
not sensitive to the upper energy bound. This result is in reasonable accord
with cLDA result of about 6 eV, which in our language corresponds to
eliminating all transitions to and from the \emph{4f} states. Due to a
slightly more extended \emph{4f} orbital, the result for $\alpha$Ce is
somewhat smaller than that of $\gamma$Ce but otherwise they are very similar.

\subsection{\emph{3d} transition metal series}%

\begin{figure}
[ptb]
\begin{center}
\includegraphics[
height=1.9804in,
width=2.7017in
]%
{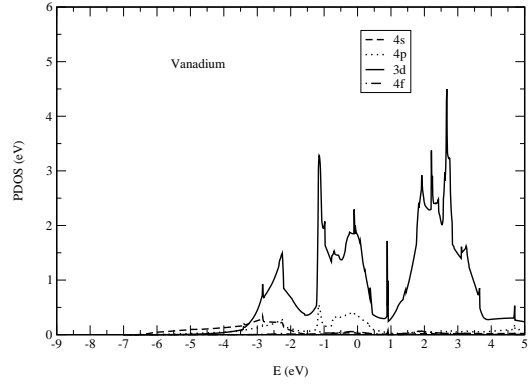}%
\caption{Partial density of states of vanadium.}%
\label{dosv}%
\end{center}
\end{figure}
%

\begin{figure}
[ptb]
\begin{center}
\includegraphics[
height=1.9899in,
width=2.7155in
]%
{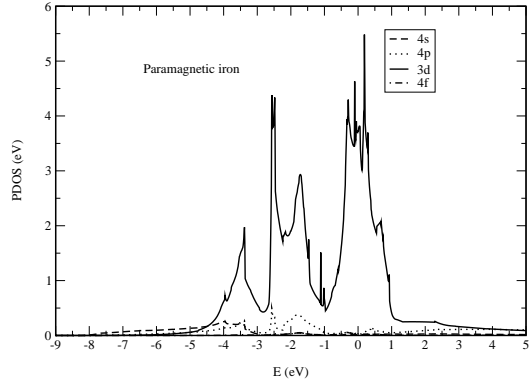}%
\caption{Partial density of states of paramagnetic iron.}%
\label{dosfe}%
\end{center}
\end{figure}
%

\begin{figure}
[ptb]
\begin{center}
\includegraphics[
height=2.0003in,
width=2.7294in
]%
{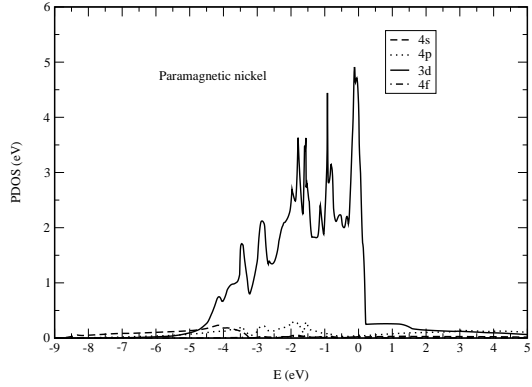}%
\caption{Partial density of states of paramagnetic nickel.}%
\label{dosni}%
\end{center}
\end{figure}

\begin{table}[ptb]%
\begin{tabular}
[c]{ll}%
Energy window (eV) & U (eV)\\
(-2.0, 4.0) & 3.7\\
(-3.0, 4.0) & 6.3\\
(-4.0, 4.0) & 7.0\\
(-2.0, 5.0) & 3.8\\
(-2.0, 6.0) & 3.9
\end{tabular}
\caption{$U$ of V as a function of energy window}%
\end{table}

\begin{table}[ptb]%
\begin{tabular}
[c]{ll}%
Energy window (eV) & U (eV)\\
(-3.0, 1.2) & 4.0\\
(-4.0, 1.2) & 4.8\\
(-5.0, 1.2) & 5.0\\
(-3.0, 2.0) & 4.3\\
(-3.0, 3.0) & 5.3
\end{tabular}
\caption{$U$ of Fe as a function of energy window}%
\end{table}

\begin{table}[ptb]%
\begin{tabular}
[c]{ll}%
Energy window (eV) & U (eV)\\
(-5.0, 0.5) & 3.7\\
(-5.0, 1.0) & 3.7\\
(-5.0, 2.0) & 6.3\\
(-6.0, 0.5) & 3.7\\
&
\end{tabular}
\caption{$U$ of Ni as a function of energy window}%
\end{table}

%

\begin{figure}
[ptb]
\begin{center}
\includegraphics[
height=2.0833in,
width=3.1445in
]%
{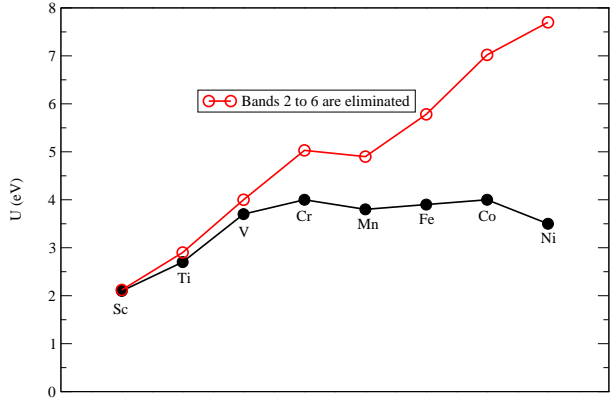}%
\caption{The Hubbard \emph{U} for the 3d series obtained by eliminating
transitions among the 3d bands. The empty circles correspond to the case where
the 3d bands are defined to be band 2 to 6 and the filled circles to the case
where the 3d bands are defined to be band number 2 up to bands below an energy
cut off above the Fermi level corresponding to a sharp drop in the 3d density
of states, as described in the text. The filled circles are what we define to
be the Hubbard \emph{U}.}%
\label{u2-6}%
\end{center}
\end{figure}

The Hubbard \emph{U} for the \emph{3d} metals is difficult to determine
unambiguously because of the strong hybridization between the \emph{3d} and
\emph{4s-4p} orbitals. Thus the result can depend strongly on the energy
window or on what we choose as our one-particle band in the Hubbard model.
This is especially true in the early transition metals, as can be seen, for
example, in Table III for vanadium. The dependence on the energy window is
less strong for the late transition metals Fe and Ni, as shown respectively in
Tables IV and V. For this reason we have adopted the following procedure for
defining the \emph{3d} band: it is defined to be those states from the second
lowest band (the lowest band is of \emph{4s} character), up to states below an
energy, just above the Fermi level, corresponding to a sharp drop in the
density of states. Figs. \ref{dosv} , \ref{dosfe}, and \ref{dosni} show the
partial density of states of V, Fe, and Ni, respectively.

One might wonder if it would not be better to project out the \emph{3d}
orbitals in calculating $P_{r}$ in (\ref{Wr}). This procedure turned out to be
mathematically unstable resulting in negative static \emph{U }and an ad hoc
procedure was employed to avoid this problem \cite{kotani}. It is also
tempting to adopt a simple procedure whereby one eliminates bands 2 to 6
(corresponding to the five "\emph{3d} states") when calculating the
polarization. This procedure is untenable because band number 6, for example,
can be very high in energy, up to 8.0 eV, which clearly cannot be regarded as
\emph{3d} states. The result using this procedure is illustrated in Fig.
\ref{u2-6}. The procedure in fact works quite well for the early elements but
the result deviates widely from that calculated using the procedure adopted in
the present work. The reason for the wide deviation is due to the increasing
number of \emph{3d} electrons as we go to the later elements. Transitions from
the occupied states to unoccupied states just above the Fermi level where the
hybridization between the \emph{3d} and \emph{4p} is strong become
increasingly numerous since the number of occupied states increases.

%

\begin{figure}
[ptb]
\begin{center}
\includegraphics[
height=2.1153in,
width=3.2344in
]%
{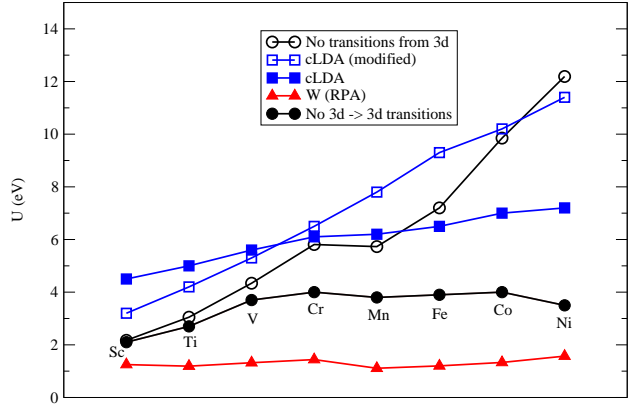}%
\caption{The Hubbard U for the 3d transition metal series calculated using the
cRPA (filled circle, the same as the filled circles in Fig. \ref{u2-6}) and
cLDA (filled square). Empty circles correspond to the cRPA result excluding
all transitions from the 3d. This should be compared with the modified cLDA
result (empty square), as described in the text. For compariosn, the fully
screened interaction \emph{W} is also plotted.}%
\label{urpa}%
\end{center}
\end{figure}

%

\begin{figure}
[ptb]
\begin{center}
\includegraphics[
height=2.0392in,
width=3.077in
]%
{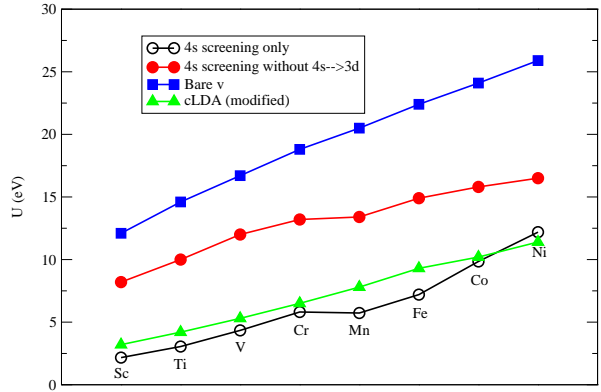}%
\caption{The effect of orbital contraction across the 3d series. Since the 3d
orbital in nickel is most localized the bare Coulomb integral (filled square)
is the largest. The bare Coulomb interaction is substantially reduced by the
4s screening (empty circle, the same as the empty circles in Fig. \ref{urpa}).
The effect of 4s to 3d screening is illustrated by the filled circles. The
effect, represented by the difference between the filled circles and the empty
circles, is slightly more important in the early than in the late elements,
due to the decreasing number of empty 3d states as we go from early to late
elements.}%
\label{u1gte}%
\end{center}
\end{figure}
In Fig. \ref{urpa} the Hubbard \emph{U} for the \emph{3d} series calculated
using the cRPA and cLDA method is shown. It is instructive to first consider
the case when all transitions \emph{from} the \emph{3d} band are eliminated
from the polarization (empty circle), i.e., all screening comes from the
\emph{4s} electrons. Note that we distinguish between transitions \emph{from}
and \emph{to} the \emph{3d} band. The screened interaction increases almost
monotonically from the early to late elements. Two factors are responsible for
this monotonic increase. Firstly, the \emph{3d} orbital contracts as we go
from early to late elements. The effect of the contraction is illustrated by
the bare Coulomb matrix element in Fig. \ref{u1gte}. Since the bare Coulomb
interaction is the same for all elements the increase must be due to the
orbital contraction. Secondly, as more \emph{3d} states are filled as we go
from early to late elements the transitions from the \emph{4s} to unoccupied
\emph{3d} states is reduced resulting in larger screened interaction. To see
the contribution of \emph{4s} to \emph{3d} transitions, the \emph{U} values
without these transitions are plotted in Fig. \ref{u1gte} (filled circle). The
curve is somewhat more flat compared with the case where \emph{4s} to
\emph{3d} transitions are included (empty circle) showing larger \emph{4s-3d}
screening in early elements. The dominant effect, however, originates from
orbital contraction. A similar conclusion was reached by Nakamura \emph{et al
}\cite{nakamura}.

The values of the Hubbard \emph{U} (filled circle) calculated within the cRPA
by eliminating \emph{3d} to \emph{3d} transitions lie between 2-4 eV across
the series. It increases from the early transition metal Sc and reaches a
maximum in the middle of the series (Cr, Mn). Compared to the case with
\emph{4s} screening only (empty circle) we have additional screening channels
arising from transitions from \emph{3d} to non-\emph{3d} bands. Evidently in
Sc, due to the small number of \emph{3d} electrons, these additional screening
channels contribute little to screening. As the number of \emph{3d }electrons
increases, these additional channels correspondingly increase their
contribution to screening reaching a maximum in nickel. In the case of nickel,
this contribution is so large that when eliminated the bare Coulomb
interaction is only reduced to about 12 eV from the bare value of 25 eV. This
is in stark contrast to Sc where the screened interaction without contribution
to screening from the \emph{3d }electrons is close to the value of the fully
screened interaction \emph{W}, implying that the small number of \emph{3d}
electrons contribute little to screening, as expected. Contrary to intuition,
the screening for \emph{U} in transition metals is not only affected by the
\emph{4s} electrons but also by the \emph{3d} electrons, especially in the
late transition elements.

We now include the remaining \emph{3d} to \emph{3d} transitions to reach the
fully screened interaction \emph{W} (triangle). Not surprisingly, the largest
contributions from \emph{3d} to \emph{3d} transitions occurs around the middle
of the series where the \emph{3d} band is half-filled while the early elements
show smaller contribution. The late elements also show tendency for smaller
\emph{3d} to \emph{3d} contribution although not as small as for the early
elements. It is remarkable although not surprising that the fully screened
interaction \emph{W} is almost constant across the series. The fact that
\emph{W} is almost constant across the series can be understood generally in
terms of metallic screening. Since metallic screening is very efficient,
essentially independent of the number of \emph{3d} electrons, the bare Coulomb
interaction is always completely screened. Unlike the case of the bare
interaction, the effect of orbital contraction is small because the static
screened interaction \emph{W} is presumably short range.

The Hubbard \emph{U} calculated using the cLDA method is significantly higher
(%
$>$%
2 eV) than the value obtained using the RPA. This is rather puzzling because
intuitively the procedure employed in the cLDA method is physically equivalent
to that of the cRPA (apart from the neglect of exchange-correlation, which is
not expected to play a big role in this case and which can be incorporated
within the LDA if necessary \cite{note1}). One possible source of difference
may be due to the fact that by constraining the number of \emph{3d }electrons
on one site in a supercell, one effectively cuts off the hopping from the
\emph{3d }orbital to other orbitals. In the language of cRPA, this
approximately corresponds to prohibiting transitions from the \emph{3d} bands
to bands other than \emph{3d }as well as transitions from non-\emph{3d} bands
to the \emph{3d} bands. For the early transition elements, the contribution of
these transitions to screening is relatively small because of the small number
of \emph{3d} electrons. For this reason cLDA result is close to that of cRPA.
However, as we approach the late elements, \emph{3d }to non-\emph{3d}
transitions contribute very significantly to screening. Since these are
neglected in cLDA, the discrepancy between cLDA\ and cRPA results becomes much
larger. This conclusion is however only semi-quantitative. We emphasize that
due to the strong hybridization between the \emph{3d} and \emph{4s-4p}
orbitals, it is difficult to make a clear comparison between cLDA and cRPA.%

\begin{figure}
[ptb]
\begin{center}
\includegraphics[
height=2.2762in,
width=3.1237in
]%
{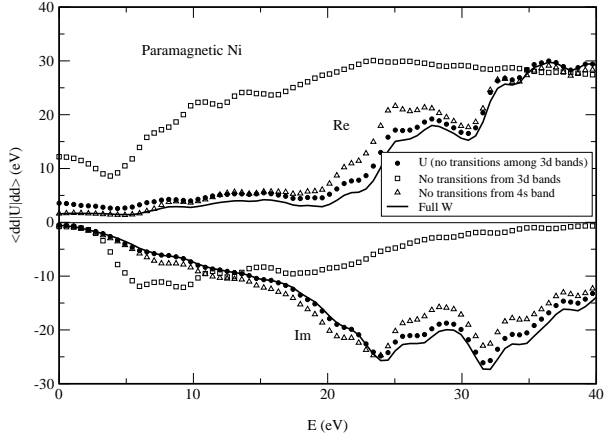}%
\caption{Frequency dependent \emph{U} (filled circle) and \emph{W} (solid
line) of paramagnetic nickel. Also shown are the results when all transitions
from the 3d bands are eliminated (square) and when all transitions from the 4s
band (lowest band) are eliminated (triangle).}%
\label{uni}%
\end{center}
\end{figure}
%

\begin{figure}
[ptb]
\begin{center}
\includegraphics[
height=2.2515in,
width=3.0909in
]%
{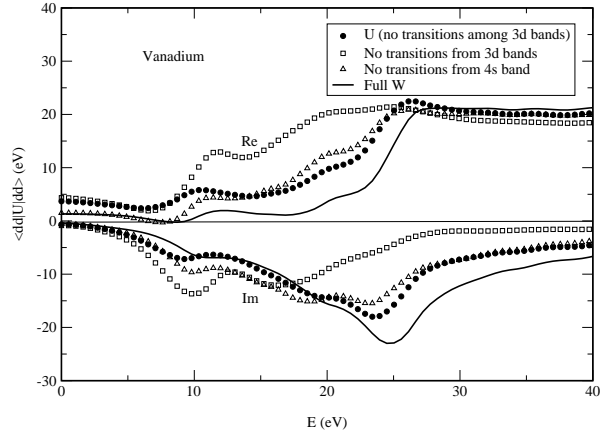}%
\caption{The same as Fig.( \ref{uni}) but for vanadium.}%
\label{v}%
\end{center}
\end{figure}

In Figs. \ref{uni} and \ref{v} we show the frequency dependence of the Hubbard
\emph{U} of nickel and vanadium. We have also analyzed the role of the
screening channels in these two systems. As already discussed before, the
difference between nickel and vanadium when all transitions from the \emph{3d}
are eliminated is very striking. More surprisingly, eliminating all
transitions from the \emph{4s} band (triangles) has little effect on the fully
screened interaction \emph{W} for both Ni and V, implying that \emph{4s}
screening is not important. On the other hand, when only the \emph{4s}
electrons are allowed to screen (by eliminating all transitions from the
\emph{3d} bands (squares)), the bare interaction is reduced from about 25 eV
to 12 eV in the case of nickel, and from 20 eV to 4 eV in the case of
vanadium, implying the importance of \emph{4s} screening. The results appear
at first sight to be contradictory but these results reflect the fact that the
screening process is \textquotedblright non-additive\textquotedblright\ in the
sense that the effect of screening depends on what is to be screened. When the
bare Coulomb interaction is already screened by the \emph{3d} electrons, the
\emph{4s} electrons have little left to screen and vice versa when the
\emph{4s} electrons have performed the screening there is not much left for
the \emph{3d} electrons to screen.

\subsection{Breathing or relaxation effects}

Finally we comment on relaxation effects of the orbitals on the cLDA method.
In calculating \emph{U} the corresponding (localized) orbitals are usually
allowed to breathe or relax in a self-consistency cycle. Constraining the
number of electrons occupying the given orbitals is physically equivalent to
applying a perturbing potential acting only on those orbitals such that the
occupation number remains constant. The relaxation of the orbitals can
therefore be regarded as the response to this perturbing potential. Since
hopping from, say, the \emph{3d} orbitals is cut off, there are no relaxation
to other non-\emph{3d} orbitals and the relaxation corresponds therefore to
(in the RPA language) transitions between \emph{3d-3d, 3d-4d}, etc.
Transitions from \emph{3d} to higher \emph{d }should be included but at least
for the case of nickel, \emph{3d-4d} \ transitions contribute little to the
screening of \emph{U}. This is in contrast to the atomic case where
\emph{3d-4d} transitions can be significant \cite{gunnarsson1}. However,
\emph{3d-3d} transitions should not be allowed since they are implicit in the
Hubbard model. The effect of \emph{3d-3d} screening can be large since this
screening is metallic and may fortuitously compensate for the missing
transitions to and from non-\emph{3d} band. For this reason, we think that
cLDA calculations for solids should not include relaxation of the constrained
orbitals, as done in the modified cLDA in the present work.

\subsection{Frequency dependence of \emph{U}}%

\begin{figure}
[ptb]
\begin{center}
\includegraphics[
height=1.8066in,
width=2.45in
]%
{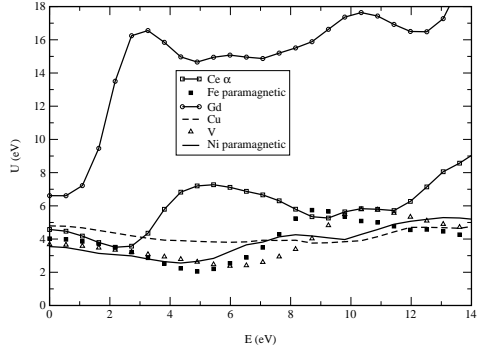}%
\caption{Frequency-dependent \emph{U} of $\alpha$Ce, Fe, Gd, Cu, V, and Ni.}%
\label{uw}%
\end{center}
\end{figure}

Finally we comment on the possible importance of energy dependence of
\emph{U}. In Fig. \ref{uw} we show the frequency dependence of \emph{U} for
some elements within a relatively low energy range. It becomes clear from the
figure that \emph{U} can be far from a constant within the band width of each
element. This suggests that taking the static value of \emph{U} and using it
in a model Hamiltonian may be inappropriate. Rather, some kind of averaging
over some range of frequency may be necessary. Consider for example cerium.
The commonly used value of 6 eV for \emph{U }in model calculations falls
within the minimum and the maximum value within a frequency range of 5 eV. For
iron, the calculated static value of about 4 eV is considerably larger than
the commonly used value of 2.0-2.5 eV, which lies in between the calculated
static value and the minimum of 2.0 eV at about $\omega$=5 eV. The case of Gd
is even more striking, with \emph{U} starting at $\sim$6.5 eV and shooting up
to $\sim$16 eV within a frequency range of 3 eV. On the other hand, nickel and
SrVO$_{3}$ appear to have a relatively constant value of \emph{U} at low energy.

\section{Summary and conclusion}

We have calculated the Hubbard \emph{U} for the \emph{3d} transition metal
series as well as cerium, SrVO$_{3}\,$and YTiO$_{3}$ using a recently proposed
basis-independent cRPA method. Although the \emph{U }values presented in this
work may not be the ultimate ones, we nevertheless believe that they provide a
useful guideline for reasonable values of \emph{U }that should be used in the
\emph{lattice} Hubbard model. The precise values of \emph{U} depends on the
choice of the one-particle orbital defining the annihilation and creation
operators of the model Hamiltonian, no matter what method we use to calculate
\emph{U}. Our scheme, however, allows the calculation of $U(\mathbf{r,r}%
^{\prime})$, which is basis-independent. But the matrix element in Eq.
(\ref{U}) entering the Hubbard model will inevitably depend on the choice of
the one-particle basis $\phi_{3d}$ of the model.

We have studied systematically the role of the screening channels in a number
of materials. In the case of cerium, \emph{5d--%
$>$%
4f} polarization is important whereas \emph{4f--%
$>$%
non-4f} is much less important due to a small number of \emph{4f} electrons.
In the case of SrVO$_{3}$, O$_{2p}$--%
$>$%
V$_{3d(t_{2g})}$ polarization plays a crucial role in reducing the Coulomb
interaction. On the other hand, t$_{2g}$--%
$>$%
e$_{g}$ polarization is not important. The screening properties of the
\emph{3d} transition series are more difficult to analyze due to strong
hybridization between the \emph{3d} and \emph{4s-4p} orbitals, which prevents
a clear distinction between the various screening channels. Nevertheless, our
results indicate that in early transition series, such as vanadium, most of
the (static) screening of \emph{U }can be attributed to the \emph{4s}
electrons. The situation is very different for the late transition metals,
where, in addition to the \emph{4s} electrons, the \emph{3d} electrons
themselves contribute substantially to the screening of \emph{U}. This can be
understood from a simple fact that the number of \emph{3d} electrons is large
so that \emph{3d--%
$>$%
}non\emph{-3d} screening is substantial, in contrast to the early transition
metals with a small number of \emph{3d} electrons. For the fully screened
interaction \emph{W}, \emph{3d--%
$>$%
3d} screening alone is sufficient to obtain the static value due to a very
efficient metallic screening for all the elements in the series. Screening
from the \emph{4s} electrons alone are not sufficient to obtain the static
\emph{W, }contrary to what is commonly stated in the literature.

We have also studied the frequency dependence of \emph{U} and found that it
can be far from constant at low energy. This suggests that for some materials
the calculated static value of \emph{U} may not be appropriate and some kind
of averaging over a frequency range may be necessary. For example, a simple
energy averaging over the band width is a possible choice.

We have also compared the cRPA values with those calculated using the
well-known cLDA\ method. Significant discrepancy is observed particularly
towards the end of the \emph{3d} series. A possible source of discrepancy may
be attributed to \emph{technical }difficulties of including polarization of
the \emph{3d} electrons to other angular momenta in the cLDA method. We would
like to emphasize that the two methods, cLDA and cRPA, theoretically ought to
give the same results. In practical implementations, \emph{3d}$\rightarrow
$non-\emph{3d} polarization corresponding to transitions from the occupied
\emph{3d} bands to unoccupied bands other than of \emph{3d} characters may not
be properly included due to the constraint on the number of the \emph{3d}
electrons, which effectively cuts off hopping of the \emph{3d} electrons to
other bands. While this polarization is small for the early \emph{3d}
elements, due to the small number of \emph{3d} electrons, the contribution
becomes increasingly large as we go towards the end of the series. Our cRPA
results for Ce and SrVO$_{3}$ are also consistent with the cLDA results but
only when the screening channels associated with the cut off of the hopping in
cLDA are correspondingly eliminated. It would be desirable to modify the
cLDA\ method to include screening channels not included in current implementations.

\section{Acknowledgment}

Stimulating and fruitful discussions with O. Gunnarsson and K. Nakamura are
gratefully acknowledged. F.A acknowledges the support from NAREGI Nanoscience
Project, Ministry of Education, Culture, Sports, Science and Technology, Japan.


\begin{thebibliography}{99}
\bibitem {note1}We have done modified cLDA calculations according to Eq.
(\ref{mod1}), using a $V(r)$ without the exchange-correlation contribution,
and the change is indeed minor (less than 0.2 eV).
\end{thebibliography}
\end{document}